\documentclass[superscriptaddress,aps,prd,preprintnumbers,nofootinbib,preprint,12pt]{revtex4}
\usepackage[utf8]{inputenc}
\usepackage[citecolor=blue,urlcolor=blue,unicode=true,pdfusetitle,
bookmarks=true,bookmarksnumbered=true,bookmarksopen=true,bookmarksopenlevel=3,
breaklinks=false,pdfborder={0 0 1},backref=false,colorlinks=true]{hyperref}
\usepackage{bm}
\usepackage{latexsym}
\usepackage{dcolumn,xcolor}
\usepackage[normalem]{ulem}
\usepackage{amsmath,amsfonts,amssymb}
\usepackage{graphicx,epsfig}
\usepackage{soul,float}
\usepackage{ulem}
\usepackage{bbold}
\usepackage{amsthm}
\usepackage{color,float} 
\usepackage{amssymb}
\usepackage{braket}
\usepackage{physics}
\usepackage{tikz}
\newcommand{\orcidicon}{%
	\begin{tikzpicture}
	\draw[lime, fill=lime] (0,0) 
	circle [radius=0.16] 
	node[white] {{\fontfamily{qag}\selectfont \tiny ID}};
	\draw[white, fill=white] (-0.0625,0.095) 
	circle [radius=0.007];
	\end{tikzpicture}	\hspace{-2mm}
}

\newcommand\orcidSebastian{{\href{https://orcid.org/0000-0001-6235-120X}{\orcidicon}}}
\newcommand\orcidJames{{\href{https://orcid.org/0000-0002-3619-2505}{\orcidicon}}}

\usepackage{color}

\begin{document}
	\title{A Quantum Informational  Approach to the Problem of Time  }
	\author{Salman Sajad Wani}
		\email{salmansajadwani@gmail.com}
		\affiliation{\scriptsize{Department of Physics Eng., Istanbul Technical University, 34469, Istanbul, Turkey}}
	\affiliation{\scriptsize{Canadian Quantum Research Center 204-3002  32 Ave Vernon, BC V1T 2L7 Canada}}

	\author{James Q. Quach\orcidJames\!\!}
		\email{quach.james@gmail.com}
		\affiliation{\scriptsize{Institute for Photonics and Advanced Sensing and School of Physical Science,
The University of Adelaide, South Australia 5005, Australia}}

	\author{Mir Faizal}
		\email{mirfaizalmir@googlemail.com}
		\affiliation{\scriptsize{Irving K. Barber School of Arts and Sciences,
University of British Columbia - Okanagan, Kelowna, British Columbia V1V 1V7, Canada}}
	\affiliation{\scriptsize{Department of Physics and Astronomy,University of Lethbridge, Lethbridge, AB T1K 3M4, Canada}}
	\affiliation{\scriptsize{Canadian Quantum Research Center 204-3002  32 Ave Vernon, BC V1T 2L7 Canada}}
	
	\author{Sebastian Bahamonde\orcidSebastian\!\!}
	\email{sbahamonde@ut.ee}
	\affiliation{\scriptsize{Laboratory of Theoretical Physics, Institute of Physics, University of Tartu, W. Ostwaldi 1, 50411 Tartu, Estonia.}}
%	\affiliation{\scriptsize{Int.Lab. Theor. Cosmology,
%			Tomsk State University of Control Systems and Radioelectronics (TUSUR), 634050 Tomsk, Russia}}

	\author{Behnam Pourhassan}
		\email{b.pourhassan@du.ac.ir}
		\affiliation{\scriptsize{Canadian Quantum Research Center 204-3002  32 Ave Vernon, BC V1T 2L7 Canada}}
	\date{\today}
	\begin{abstract}
Several novel approaches have been proposed to resolve the problem of time by relating it to change. We argue using quantum information theory that the Hamiltonian  constraint in quantum gravity cannot probe change, so it cannot be used to obtain a meaningful notion  of time. This is due to the  absence of quantum Fisher information with respect to the  quantum  Hamiltonian of  a time-reparametization invariant  system. We also observe that the inability of this   Hamiltonian to probe change can be related to its inability to  discriminate between states of such a system.   
However, if the  time-reparametization symmetry is spontaneously broken due to the formation  of  quantum cosmological time crystals,  these problems can be resolved, and it is possible  for time to emerge in quantum gravity. 
	\end{abstract}
	\maketitle

Diffeomorphism invariance can be viewed as  a gauge degree of freedom, so the Hamiltonian in quantum gravity is only a generator of a gauge transformation. Thus, it is not possible to directly define a meaningful notion of   time  in quantum gravity, and this absence of time is known as the problem of time.  This problem occurs in most approaches to quantum gravity \cite{Arnowitt:1959ah, Henneaux:2018cst, Okolow:2013lwa, Peldan:1991as,Carlip:2018zsk, Halliwell:1988wc, Bojowald:2006zi, Smolin:2010iq, Boyarsky:2004bu, Gusin:2008zz, Rodrigo:1985km, Rodrigo:1984mj, Hamber:2011cn, Baytas:2016cbs, Zhang:2011vi, Alonso-Serrano:2018zpi, Baratin:2010wi, Gielen:2013kla}.
As the notion of time is central to cosmology, one needs to resolve the problem of time in quantum gravity to construct meaningful model of quantum cosmology. Several different novel proposals have been suggested to resolve the problem of time, such as  the  frozen formalism \cite{Anderson:2012vk}, conditional probabilities \cite{Gambini:2008ke},  use of scale factor as time \cite{DiTucci:2019bui},  matrix formulation of quantum gravity \cite{Carlini:1994fp}, use of suitable operators for Cauchy surfaces \cite{Wald:1993kj}, unimodular gravity in the  Ashtekar formulations \cite{Smolin:2010iq}, the use of matter fields as time  \cite{Husain:2011tk},   rigging map of group averaging \cite{Marolf:2009wp},  non-perturbative quantum gravity \cite{Husain:2015dxa, Rovelli:1993bm}, third quantization \cite{Ohkuwa:2012cm, Campanelli:2020swy} and even  fourth quantization \cite{Faizal:2013soa}.

These proposals are based on the  notion that time is associated with   change in the quantum states of the system. To observe such a change, one needs to  probe the state by an Hermitian operator, which can only be done if that operator  contains  information about such states. So, it is important to analyse  quantum information associated with the Hamiltonian operator in quantum cosmology. 
In the  minisuperspace approximation, the Hamiltonian  in several  quantum  cosmological systems, mathematically resembles the Hamiltonian of simple time-reparametrization invariant quantum mechanical systems  \cite{Guimarey:2019lmn, Janssen:2019sex, Djordjevic:2018mzs, Chagoya:2016gcm}. As quantum  cosmological models can be difficult to interpret conceptually, we will instead analyze a simple quantum mechanical system with  time-reparametrization invariance, without loss of applicability. 

The amount of information that an operator contains about the states of a system can be quantified  by the quantum Fisher information (QFI)  \cite{Braunstein:1994zz, toth2013extremal, paris2009quantum, genoni2008optimal}, which for a  Hamiltonian ${H}$  with respect to a density matrix  ${\rho}$  is~\cite{Braunstein:1994zz, toth2013extremal, paris2009quantum, genoni2008optimal} 
  \begin{equation}
   F[{\rho},{H}]=\text{Tr}[H,[{\rho}, {L}]]\,, 
   \end{equation}
where    ${L}$ is  the symmetric logarithmic derivative~\cite{Braunstein:1994zz, toth2013extremal, paris2009quantum, genoni2008optimal},  
\begin{equation}
{L}=2 i \sum_{i,l}\frac{\langle H_{l}|[{H}, {\rho} ]| H_{i}\rangle}{\varepsilon_i+\varepsilon_l}|H_{i}\rangle\langle H_{l}|\, , 
\end{equation}
with  $\varepsilon_l$ as the eigenvalues of   ${\rho}$. Due to time-reparametrization invariance  all the different eigenstates of 
${H}$, represented here by $ |H_{l}\rangle$, are related by a gauge transformation, and so they are  physically equivalent with  the same eigenvalue $\epsilon$.  Hence, we have    $[{H}, {\rho} ] =0$, and the QFI of the Hamiltonian for such systems vanish,  
\begin{equation}
     F [{\rho},{H}]= 0\,.
 \end{equation}
 
As  the Hamiltonian does not contain any QFI about the states  of a system, it cannot probe  changes in those states. This can be explicitly be shown using quantum state discrimination  \cite{cook2007optical, bae2015quantum, becerra2013implementation, leverrier2009unconditional}.  
Let Alice send  probabilities  $\{p_i\} $ associated with states of a  time-reparametrization system   $\{{\rho}_i\} $ to Bob through a classical channel. She then sends a  prepared quantum state through a quantum channel.  Even though  there will be an error in identifying  this state, Bob can optimize the odds of correctly identifying it using minimum error discrimination \cite{cook2007optical, bae2015quantum, becerra2013implementation, leverrier2009unconditional}. 
Let the  outcome  $\epsilon_i $,  associated with positive-operator-valued measure  (POVM) ${\pi_i} $, be used to  indicate  ${\rho}_i = |\psi_i\rangle\langle\psi_i| $, then  we can minimize  the error in  discrimination of the non-degenerate set $\{ \rho_i \}$ by  choosing a set of POVMs which  minimizes the expression \cite{han2020helstrom,  flatt2019multiple}
\begin{eqnarray}
    P_{\rm err}&=&\sum_{i=1}^{n}p_i\sum_{j\neq i}\text{Tr}({\rho_i} {\pi_j})\,,
    \nonumber\\
    &=& p_1\Big(P(\epsilon_2|\psi_1) +  P(\epsilon_3|\psi_1)+\ldots+P(\epsilon_{n}|\psi_{1})\Big)+\nonumber\\
    && \ldots+p_n\Big(P(\epsilon_1|\psi_n) + P(\epsilon_2|\psi_n)+\ldots+P(\epsilon_{n-1}|\psi_{n})\Big)\,,
\end{eqnarray}
where $P(\epsilon_i|\psi_j)$ is the probability for the outcome to be $ \epsilon_i$, if the state is $|\psi_j\rangle$. In the presence of degeneracy, the expression is modified to  $P_{\rm err}=  \sum_{i=1}^{n} p_i\sum_{j\neq i} N_j \text{Tr}({\rho_i} {\pi_j})$, where normalization constants $ N_j=1/m$ (with $j = 1\ldots m$)  are used to account for $m$ degenerate states. As the Hamiltonian is a constraint with a fixed eigenvalue $\epsilon$, $ \pi_i = \pi_j $ and $N_j =1/n~,\forall\, i,j$,  the probability in making an error in discriminating the state is   
\begin{eqnarray}
    P_{\rm err}
    &=& \frac{1}{n}\Big[p_1\Big(P(\epsilon|\psi_1) +  P(\epsilon|\psi_1)+\ldots+P(\epsilon_{n}|\psi_{1})\Big)+\nonumber\\
    && \ldots+p_n\Big(P(\epsilon|\psi_n) + P(\epsilon|\psi_n)+\ldots+P(\epsilon|\psi_{n})\Big) \Big]\,,
    \nonumber \\
    &=& \frac{n -1}{n} \sum_{i = 1}^{n} p_i =  \frac{n -1}{n}\,, 
\end{eqnarray}
where we have used $P(\epsilon |\psi_{j})=1$ and $\sum_{i = 1}^{n} p_i = 1$. This error  $1-1/n$ is exactly equal to the error obtained by randomly guessing the outcome; in other words,  Bob cannot use the Hamiltonian operator to discriminate  the states sent by Alice. 
\begin{figure}[tb]
    \centering
    \includegraphics[width=\columnwidth]{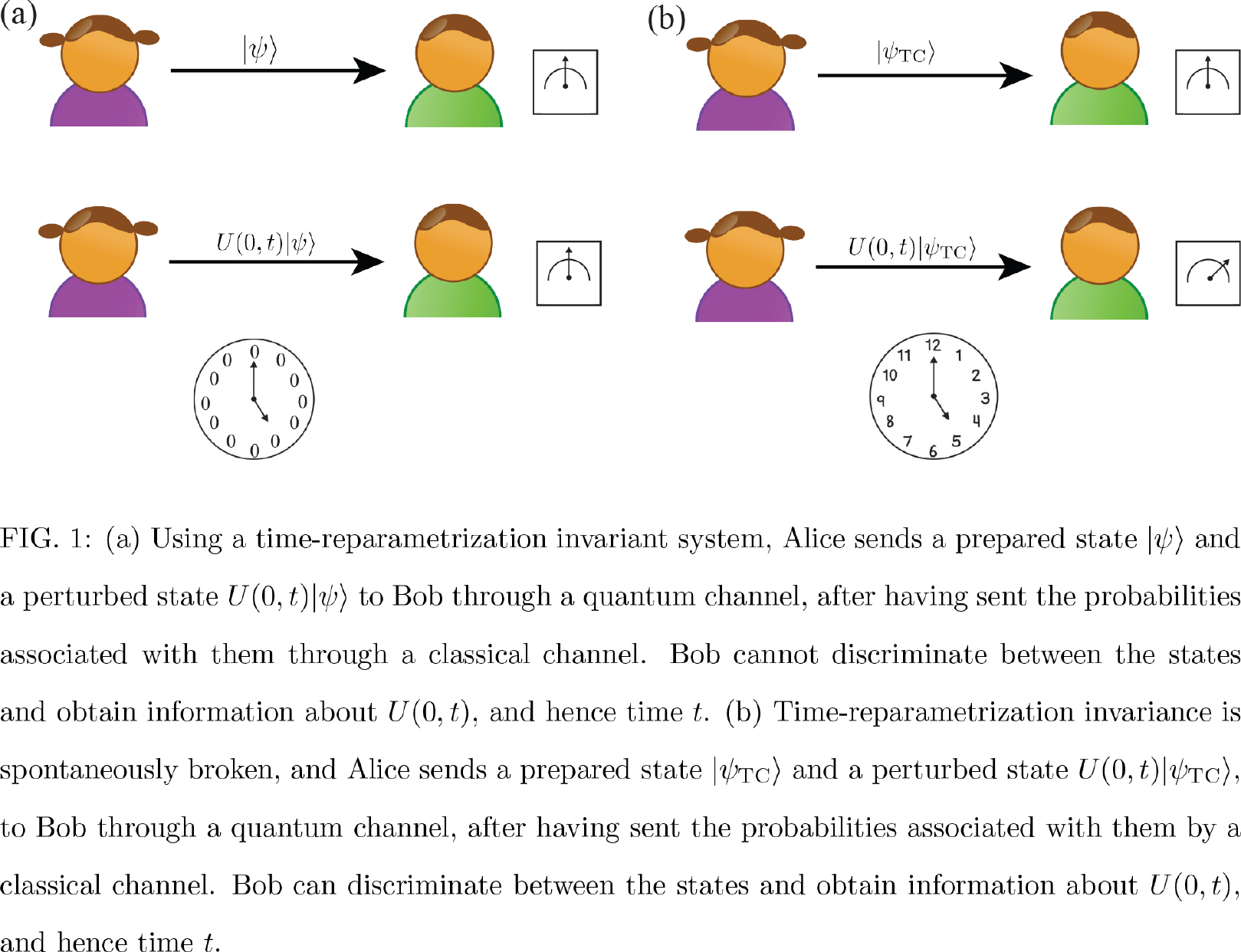}
    %\caption{(a) Using a time-reparametrization invariant system, Alice sends a prepared state $|\psi\rangle$ and a perturbed state $U(0,t)|\psi\rangle$  to Bob through a quantum channel, after having sent the probabilities associated with them through a classical channel. Bob cannot discriminate between the states and obtain information about $U(0, t)$, and hence time $t$. 
    %(b) Time-reparametrization invariance is spontaneously broken, and Alice sends a prepared state $|\psi_\text{TC}\rangle$ and a perturbed state $U(0,t)|\psi_\text{TC}\rangle$, to Bob through a quantum channel, after having sent the probabilities associated with them by a classical channel. Bob can discriminate between the states and obtain information about $U(0, t)$, and hence time $t$. }
    \label{fig:my_label}
\end{figure}
    
An important consequence of the inability to discriminate between states is that one can not observe changes in states, and therefore have no notion of time flow.  This is explicitly demonstrated if we let  Alice take the same ensemble of states, and perturbs them over a fixed amount of time with   operator ${U}(0,t)$ (such that  $ |\psi_{i}(t)\rangle  = {U}|\psi_i(0)\rangle  $). Alice again sends  the  probabilities  $\{p_{i}(t)\} $ associated with the perturbed states   $\{{\rho}_{i}(t)\} $ through  a classical channel, and a prepared state   through   a quantum channel to Bob. Repeating the previous argument, we  observe that the probability of   error in discriminating these perturbed  states is again  $P_{\rm err}=1-1/n$.  Bob cannot measure the change produced by the perturbation $U(0, t)$, so he cannot know about the time interval $t$.  This is depicted in Fig.~1a. 

As the absence of information in the Hamiltonian is due to its time-reparametrization invariance, it seems that this problem could be resolved by breaking it. This could be done by modifying the Einstein-Hilbert action; however,  this results in theoretical and observational problems at the classical level~\cite{Rovelli:1990ph}. A better resolution is to retain the time-reparametrization invariance of   the classical Einstein-Hilbert action, and propose that  the   quantum gravitational vacuum state is not invariant under it. Such spontaneous symmetry  breaking of time-reparametrization invariance will produce a discrete structure in time;   in analogy with ordinary spatial crystals, this discrete temporal structure is called a time crystal   \cite{Wilczek:2012jt}. It has also been demonstrated that time crystals can form at Planck time due to quantum gravitational effects \cite{Faizal:2014mba, Khorasani:2017ffr}. The universe evolves by taking discrete steps in quantum cosmological time crystals  \cite{Faizal:2014nwa, Garattini:2015aca}.

Even though certain no-go theorems exists for   time crystals~\cite{Watanabe:2014hea}, their applicability are restrictive, and we argue that they do not apply to quantum cosmological time crystals. Firstly, the no-go theorems only hold for closed-systems; however,   brane world models can be viewed as  open systems, due to the exchange of gravitons  with the bulk  \cite{ArkaniHamed:1998rs, Giudice:1998ck, Dvali:1998pa, Hewett:1998sn}. Furthermore, even conventional cosmological models can be considered as open systems, as   Planck scale quantum fluctuations can  change the  topology of the universe,  producing  baby universes \cite{Balasubramanian:2020jhl, Marolf:2020xie};  with  the information   flowing into these baby universes getting  disconnected from the original universe.   Secondly, and perhaps more importantly, it has  been recently shown that the no-go theorems do not apply in the presence of long range forces \cite{Kozin:2019qpc}. As gravity is a long range force, these no-go theorems should not apply to quantum cosmological time crystals. 

Here we do not care about the specific details of the different quantum cosmological time crystals \cite{Faizal:2014nwa, Garattini:2015aca}. Instead we analyze the information theoretic implications of the universal feature of these models, which is the spontaneous breaking of time-reparametrization invariance, on the problem of time. After breaking  time-reparametrization invariance, the different eigenstates of the Hamiltonian are not related to each other by  gauge transformations, and  can have different eigenvalues.  So, they will not commute with an arbitrary density matrix, $[{H}, {\rho} ] \neq0$, and  the QFI will not  vanish   
\begin{equation}
      F [{\rho},{H}]\neq 0\,.
 \end{equation}  
We can repeat the previous analysis, and investigate change in such states. Let Alice again send probabilities  $\{p_{\text{TC},i}\} $  associated with states of a system, with spontaneously broken time-reparametrization invariance,  $\{{\rho}_{\text{TC},i}\} $, through  a classical channel; she sends a prepared  state though a quantum channel (the TC subscript is used to indicate states in a time-crystal background).   As the Hamiltonian can have different eigenvalues, $\pi_i \neq \pi_j$, and  $P(\epsilon_i|\psi_{\text{TC},i}) < 1, P(\epsilon_j|\psi_{\text{TC},j})< 1$.  In general,  the probability of error in discriminating between these states would be less than the error produced by randomly guessing: $P_\text{err}< 1-1/n$ \cite{cook2007optical, bae2015quantum, becerra2013implementation, leverrier2009unconditional}.   Alice takes another similar ensemble of states, and again perturbs it for a fixed amount of time by $U(0, t)$. Alice sends the  probabilities associated with the perturbed states by a classical  channel, and the perturbed states through a quantum channel.  The probability of  error  in discriminating these perturbed states is again $P_{\rm err}< 1-1/n$. As Bob can discriminate between the original states, and the perturbed states, he can also probe the change in those states represented by $U(0, t)$. Thus, Bob can know about the passage of time $t$. The breaking of time-reparametrization invariance leads to an emergence of time in the system. This is depicted in Fig~1b.

It was observed that the problem of time is intrinsic to quantum gravity, as the Hamiltonian constraint did not contain any QFI. This led to an absence of a meaningful  notion of time due to the  inability of the Hamiltonian constraint to discriminate between states. However, these problems are naturally resolved when  time-reparametrization invariance is spontaneously broken.

\section*{Acknowledgements}
S.B. was supported by the Estonian Research Council grants PRG356 ``Gauge Gravity"  and by the European Regional Development Fund through the Center of Excellence TK133 ``The Dark Side of the Universe". J.Q.Q. acknowledges the ﬁnancial support of the Ramsay Fellowship. 
 
\bibliographystyle{utphys}
\bibliography{references}

\end{document}